\documentclass[%
reprint,
twocolumn,
superscriptaddress,
showpacs,preprintnumbers,
 amsmath,amssymb,
 aps,
]{revtex4-1}

\usepackage{graphicx}% Include figure files
\usepackage{dcolumn}% Align table columns on decimal point
\usepackage{bm}% bold math

\begin{document}

%\preprint{}

\title{Temperature and composition phase diagram in the iron-based ladder compounds Ba$_{1-x} $Cs$_x$Fe$_2$Se$_3$}

\author{Takafumi Hawai}
\affiliation{Neutron Science Laboratory, Institute for Solid State Physics, University of Tokyo, Tokai, Ibaraki 319-1106, Japan}
\author{Yusuke Nambu}
\affiliation{Institute of Multidisciplinary Research for Advanced Materials, Tohoku University, Sendai 980-8577, Japan}
\author{Kenya Ohgushi}
\affiliation{Institute for Solid State Physics, University of Tokyo, Kashiwa, Chiba 277-8581, Japan}
\affiliation{Department of Physics, Tohoku University, Sendai, Miyagi 980-8578, Japan}
\author{Fei Du}
\affiliation{Institute for Solid State Physics, University of Tokyo, Kashiwa, Chiba 277-8581, Japan}
\affiliation{Key Laboratory of Physics and Technology for Advanced Batteries (Ministry of Education), College of physics, Jilin University, Changchun, 130012, People’s Republic of China}
\author{Yasuyuki Hirata}
\affiliation{Institute for Solid State Physics, University of Tokyo, Kashiwa, Chiba 277-8581, Japan}
\author{Maxim Avdeev}
\affiliation{Bragg Institute, Australian Nuclear Science and Technology Organization, Locked Bag 2001, Kirrawee DC NSW 2232, Australia}
\author{Yoshiya Uwatoko}
\affiliation{Institute for Solid State Physics, University of Tokyo, Kashiwa, Chiba 277-8581, Japan}
\author{Yurina Sekine}
\affiliation{Quantum Beam Science Directorate, Japan Atomic Energy Agency, Tokai, Ibaraki 319-1195, Japan}
\author{Hiroshi Fukazawa}
\affiliation{Quantum Beam Science Directorate, Japan Atomic Energy Agency, Tokai, Ibaraki 319-1195, Japan}
\author{Jie Ma}
\affiliation{Quantum Condensed Matter Division, Oak Ridge National Laboratory, Oak Ridge, Tennessee 37831, USA}
\author{Songxue Chi}
\affiliation{Quantum Condensed Matter Division, Oak Ridge National Laboratory, Oak Ridge, Tennessee 37831, USA}
\author{Yutaka Ueda}
\affiliation{Institute for Solid State Physics, University of Tokyo, Kashiwa, Chiba 277-8581, Japan}
\affiliation{Toyota Physical and Chemical Research Institute, Nagakute, Aichi 480-1192, Japan}
\author{Hideki Yoshizawa}
\affiliation{Neutron Science Laboratory, Institute for Solid State Physics, University of Tokyo, Tokai, Ibaraki 319-1106, Japan}
\author{Taku J. Sato}
\affiliation{Institute of Multidisciplinary Research for Advanced Materials, Tohoku University, Sendai 980-8577, Japan}

\date{\today}% It is always \today, today,

\begin{abstract}
We investigated the iron-based ladder compounds (Ba,Cs)Fe$_2$Se$_3$. Their parent compounds, BaFe$_2$Se$_3$ and CsFe$_2$Se$_3$, have different space groups, formal valences of Fe and magnetic structures.
Electrical resistivity, specific heat, magnetic susceptibility, X-ray diffraction and powder neutron diffraction measurements were conducted to obtain temperature and composition phase diagram of this system.
Block magnetism observed in BaFe$_2$Se$_3$ is drastically suppressed with Cs doping.
In contrast, stripe magnetism observed in CsFe$_2$Se$_3$ is not so fragile against Ba doping.
New type of magnetic structure appears in intermediate compositions, which is similar to stripe magnetism of CsFe$_2$Se$_3$, but inter-ladder spin configuration is different.
Intermediate compounds show insulating behavior, nevertheless finite $T$-linear contribution in specific heat was obtained at low temperatures.

\end{abstract}

\pacs{72.20.-i, 75.25.-j, 75.40.-s, 75.50.Ee}

\maketitle

%Introduction
\section{\label{introduction}INTRODUCTION}
Iron-based superconductors (SCs) have created considerable interests in condensed matter physics community since its discovery \cite{doi:10.1021/ja800073m}.
Despite intensive efforts, the superconducting mechanism is still controversial; spin fluctuation \cite{PhysRevLett.101.087004}, orbital fluctuation \cite{PhysRevLett.104.157001}, or both of them are prime candidates for mechanism of the iron-based SCs.
As these fluctuations are sensitive to crystal and magnetic structure, it is desired to explore superconductivity in iron-based compounds with different crystal and magnetic structures.

To date iron-based SCs are known to have the two dimensional Fe square lattice with Fe atoms being tetrahedrally coordinated by pnictogens or chalcogens.
They show metallic behavior in electrical resistivity, and have antiferromagnetic order\cite{0953-8984-22-20-203203, dai2012magnetism}.
The 1111 \cite{luetkens2009electronic, zhao2008structural}, 111 \cite{Chu2009326, PhysRevB.80.020504} and 122 \cite{PhysRevLett.101.257003, PhysRevB.78.140504} systems have single-stripe magnetic structure and their magnetic moments are less than 1 $ \mu_{\rm B} $.
The 11 systems have double stripe magnetic structure with $ \sim$ 2 $\mu_{\rm B} $ moment \cite{PhysRevLett.102.247001, PhysRevLett.106.057004}.
In contrast, the 245 systems, $A_2$Fe$_4$Se$_5$ ($A$ = K, Rb, and Cs), are insulators and have two-dimensional (2D) iron square lattice with periodic vacancy of Fe atoms \cite{PhysRevB.82.180520, PhysRevLett.107.137003}.
They have block magnetism with magnetic moments $ \sim$ 3 $\mu_{\rm B} $ and have been considered to be the third type of magnetic structures for parent compounds of iron-based SCs.

To gain further insights into the interplay between crystal structure, magnetism and superconductivity, research on compounds with different dimensionality is important.
The spatial dimensionality should have effects on magnetic interaction, orbital selectivity and topology of Fermi-surface of Fe atoms.
In this regard, it is noteworthy that one-dimensional spin-ladder systems (La, Y, Sr, Ca)$_{14}$Cu$_{24}$O$_{41}$ provide great insight into the understanding of the two-dimensional copper-oxide superconductors \cite{Vuletic2006169}.

Recently, iron-based ladder compounds $A$Fe$_2$$X_3$ ($A$ = K, Rb, Cs or Ba, and $X$ = chalcogen) \cite{Hong197293, Klepp19961, Mitchell20041867} have been under intensive investigation \cite{PhysRevB.85.064413, 0953-8984-25-31-315403, PhysRevB.84.245132, PhysRevB.84.180409, PhysRevB.84.214511, PhysRevB.85.214436, PhysRevB.85.180405, 0953-8984-26-2-026002, RevModPhys.85.849}.
The $A$Fe$_2$$X_3$ compounds consist of Fe$X_4$ tetrahedra, which is similar to the 2D iron-based SCs, but Fe atoms form two-leg ladder lattice giving rise to the quasi-one dimensionality.
They also show related magnetism with iron-based SCs, stripe and block magnetism.
Unlike the most iron-based SCs, iron-based ladder compounds are insulators.
KFe$_2$Se$_3$ \cite{PhysRevB.85.180405} and CsFe$_2$Se$_3$ \cite{PhysRevB.85.214436} have stripe magnetism; magnetic moments are arranged to form ferromagnetic units along the rung direction, and the units stack antiferromagnetically along the leg direction.
BaFe$_2$Se$_3$ \cite{PhysRevB.85.064413, PhysRevB.84.180409} has block magnetism; magnetic moments are arranged to form Fe$_4$ ferromagnetic units, and each units stack antiferromagnetically along the leg direction.
Interestingly, in BaFe$_2$Se$_3$, Fe lattice distortion under N$\acute{\rm e} $el temperature ($T_{\rm{N}} $) is reported \cite{PhysRevB.85.064413, PhysRevB.84.180409}.
This indicates that lattice and magnetism are strongly coupled.

Recently magnetization measurement for the iron-based ladder compounds Ba$_{1-x}$K$_x$Fe$_2$Se$_3$ have been reported \cite{PhysRevB.85.180405}.
BaFe$_2$Se$_3$ has the $Pnma$ space group, formal valence of Fe$^{2+}$ and block magnetism with $T_{{\rm N}}$ $\sim 255$ K \cite{PhysRevB.85.064413, PhysRevB.84.180409}; on the other hand KFe$_2$Se$_3$ has the $Cmcm$ space group, formal valence of Fe$^{2.5+}$ and stripe magnetism with $T_{{\rm N}}$ $\sim 200$ K \cite{PhysRevB.85.180405}.
The previous study determined $T_{{\rm N}}$ of Ba$_{1-x}$K$_x$Fe$_2$Se$_3$ from magnetization measurement and showed reduced $T_{{\rm N}}$ by K doping decreases \cite{PhysRevB.85.180405}.
However, little is known on the variation of the crystal and magnetic structures in intermediate compositions.

In this study, we focus on the mixed system of BaFe$_2$Se$_3$ and CsFe$_2$Se$_3$.
CsFe$_2$Se$_3$ has the $Cmcm$ space group, formal valence of Fe$^{2.5+}$, stripe magnetism similar to KFe$_2$Se$_3$ with $T_{{\rm N}}$  $\sim 175$ K \cite{PhysRevB.85.214436}.
Since CsFe$_2$Se$_3$ has lower $T_{{\rm N}}$ than KFe$_2$Se$_3$, we assumed that $T_{{\rm N}}$ was strongly suppressed by dilution.
We examined Ba$_{1-x}$Cs$_x$Fe$_2$Se$_3$ through electrical resistivity, specific heat, magnetic susceptibility, X-ray diffraction and powder neutron diffraction measurements, and established  composition and temperature phase diagram of Ba$_{1-x}$Cs$_x$Fe$_2$Se$_3$.
The boundary of the two distinct space groups, $Pnma$ and $Cmcm$, is between $x = 0.05$ and $0.15$.
Block magnetism in BaFe$_2$Se$_3$ is drastically suppressed by Cs doping, whereas stripe magnetism in CsFe$_2$Se$_3$ is relatively stable.
In addition, new magnetic structure is observed for $x = 0.5, 0.55$ and $ 0.65$ at the lowest temperatures.
Surprisingly at $x = 0.25$, no magnetic reflections were observed down to 7 K in the powder neutron diffraction.

\section{\label{experimental}EXPERIMENTAL}

Polycrystalline samples of Ba$_{1-x} $Cs$_x$Fe$_2$Se$_3$ ($x = 0.05, 0.15, 0.25, 0.4, 0.5, 0.55, 0.65$ and $0.75$) were synthesized from the mixture of BaFe$_2$Se$_3$ and CsFe$_2$Se$_3$.
The values of $x$ were defined as  nominal composition.
The two parent compounds were synthesized from the stoichiometric ratio of the starting materials such as Ba : Fe : Se = 1 : 2 : 3 and Cs$_2$Se : Fe : Se = 1 : 4 : 5 \cite{PhysRevB.85.064413, PhysRevB.85.214436}.
They were mixed in a carbon crucible sealed into a quartz tube with 0.3 atm of Ar gas and reacted at 900 $^\circ$C for 24 hours.

Single crystals of Ba$_{1-x} $Cs$_x$Fe$_2$Se$_3$ ($x = 0.05, 0.15, 0.25, 0.4, 0.5, 0.55, 0.65$ and $0.75$) were grown from the melt.
Stoichiometric mixtures of the two parent compounds were put in a carbon crucible and then sealed into a quartz tube with 0.3 atm of Ar gas.
They were reacted at 1100 $^\circ$C for 24 hours and cooled to 750 $^\circ$C with a rate of 6 $^\circ$C/h, and then cooled down to room temperature for 12 hours.
Reflecting one dimensionality of the structure, the crystals preferentially grew along the leg direction.

Quality of the samples was checked by the powder x-ray diffraction (XRD) with Cu-$K_{\alpha}$ radiation at room temperature.
No detectable amount of impurities were observed in the XRD measurements. 
Single-crystal XRD with Mo-$K_{\alpha}$ radiation was performed by using R-AXIS RAPID II (Rigaku) at room temperature to determine the space group of  Ba$_{1-x}$Cs$_x$Fe$_2$Se$_3$.

Electrical resistivity is measured using the four-probe method in the temperature ($T$) range between 30 and 350 K. 
Specific heat measurements were performed using relaxation method between $T$ = 1.8 and 300 K.
These measurements were performed using the Physical Properties Measurement System (Quantum Design, PPMS).

DC susceptibility was measured between $T$ = 2 and 300 K under magnetic field up to $H$ = 5 T using a commercial SQUID magnetometer (Quantum Design, MPMS).

Powder neutron diffraction was performed using the high-resolution ECHIDNA diffractometer at the Australian Nuclear Science and Technology Organization, and the Wide-Angle Neutron Diffractometer (WAND) at Oak Ridge National Laboratory. 
Experimental conditions are summarized in Table \ref{tab:table1}.
The Rietveld refinements were performed using the Fullprof suite \cite{rodriguez1993recent}.
The crystal structures were depicted by the VESTA software \cite{Momma:db5098}.

\begin{table}[t!]
\caption{Detailed diffractometer parameters. Samples are packed into vanadium cans. Temperature is controlled by using a Closed-cycle refrigerator (CCR).}
\label{tab:table1}
\begin{ruledtabular}
\begin{tabular}{ccc}
Diffractometer & ECHIDNA & WAND \\
\hline
Wave length (\AA ) & 2.4395(2) & 1.4827(1) \\
2$\theta $ range (deg.) & 2.75-163.875 & 5.025-129.625 \\
2$\theta $ step (deg.) & 0.125 & 0.2 \\
Monochromator & Ge(335) & Ge(113) \\
Refrigerator & CCR (3-300 K) & CCR (5-300 K) \\
\end{tabular}
\end{ruledtabular}
\end{table}

\section{\label{results}RESULTS}
\subsection{\label{crystal_structure}CRYSTAL STRUCTURE}
\begin{center}
\begin{figure}[htb]
\includegraphics[width=1\linewidth]{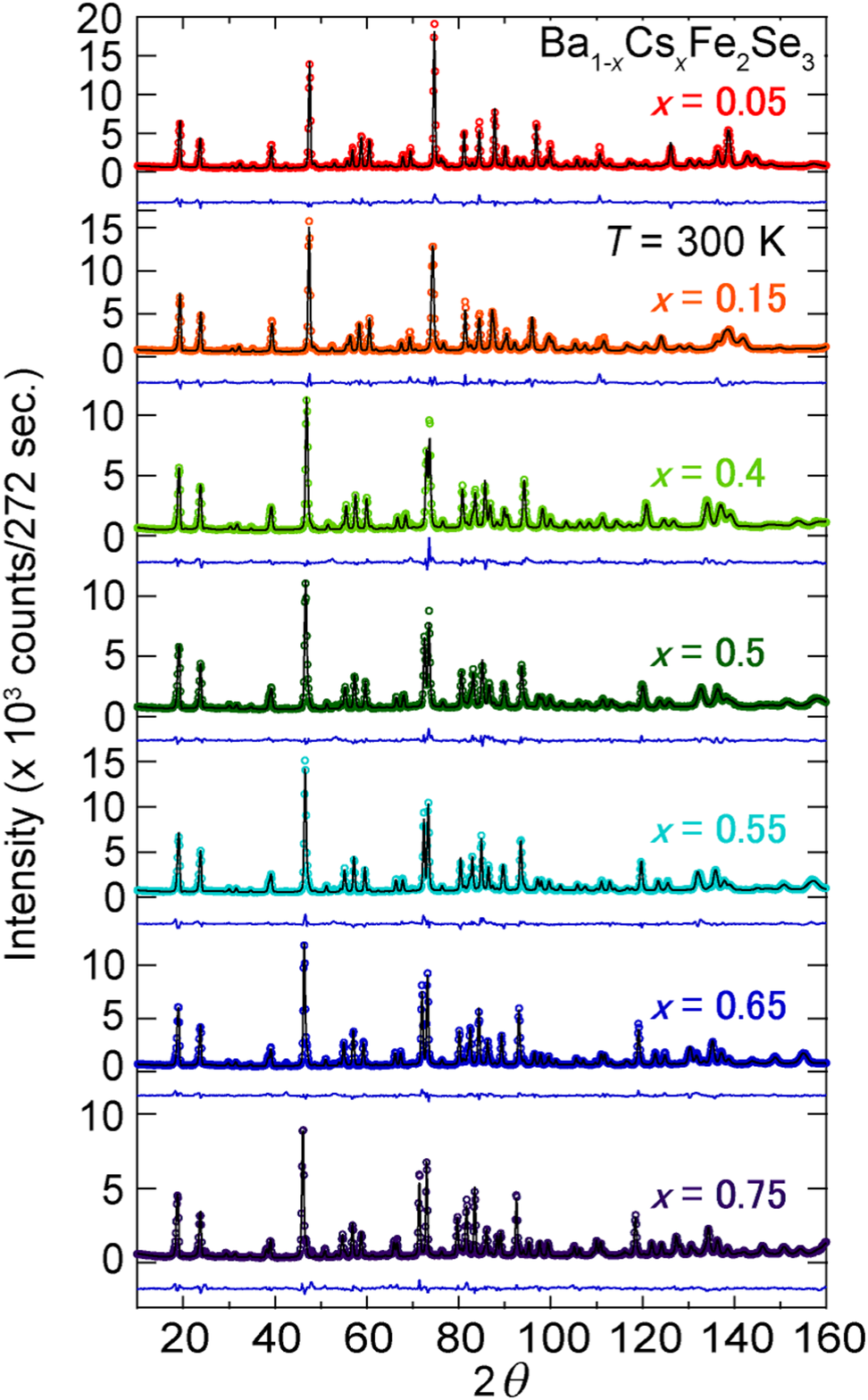}
\caption{(color online). High-resolution powder neutron diffraction patterns of Ba$_{1-x} $Cs$_x$Fe$_2$Se$_3$ at 300 K on ECHIDNA together with Rietveld analysis. The bottom blue lines give the difference between the observed (open circles) and calculated (line) intensities.}
\label{results_of_R}
\end{figure}
\end{center}

Figure \ref{results_of_R} shows neutron diffraction patterns for Ba$_{1-x} $Cs$_x$Fe$_2$Se$_3$ together with Rietveld refinement results at 300 K.
Table \ref{tab:table2} summarizes refined atomic positions, and other crystallographic data from the Rietveld refinement are shown in Table \ref{tab:table3}.
\begin{table}[htb]
\caption{\label{tab:table2}Refined atomic positions at 300 K for Ba$_{1-x} $Cs$_x$Fe$_2$Se$_3$ ($x$ = 0.05, 0.15, 0.4, 0.5, 0.55, 0.65, and 0.75). Note that $x = 0.05$ belongs to the $Pnma$ space group. The others belong to $Cmcm$.}
\begin{ruledtabular}
\begin{tabular}{cccccc}
Atom & site & $x$ & $y$ & $z$ & $U_{ \rm{iso}}$ (\AA$^2$) \\
\hline
$x =$ 0.05 \\
Ba/Cs & 4$c$ & 0.1798(8) & 1/4 & 0.508(2) & 0.020(3) \\
Fe & 8$d$ & 0.4960(4) & 0.0008(10) & 0.3544(4) & 0.0120(9) \\
Se1 & 4$c$ & 0.3632(6) & 1/4 & 0.2265(8) & 0.022(3) \\
Se2 & 4$c$ & 0.6281(6) & 1/4 & 0.4959(11) & 0.0131(19) \\
Se3 & 4$c$ & 0.3938(6) & 1/4 & 0.8100(7) & 0.007(2) \\
\hline
$x =$ 0.15 \\
Ba/Cs & 4$c$ & 1/2 & 0.1788(8) & 1/4 & 0.011(3) \\
Fe & 8$e$ & 0.3518(5) & 1/2 & 0 & 0.0223(11) \\
Se1 & 4$c$ & 0 & 0.1245(6) & 1/4 & 0.011(2) \\
Se2 & 8$g$ & 0.2073(6) & 0.3804(5) & 1/4 & 0.040(2) \\
\hline
$x =$ 0.4 \\
Ba/Cs & 4$c$ & 1/2 & 0.1725(8) & 1/4 & 0.025(3) \\
Fe & 8$e$ & 0.3527(5) & 1/2 & 0 & 0.0185(10) \\
Se1 & 4$c$ & 0 & 0.1215(5) & 1/4 & 0.0099(19) \\
Se2 & 8$g$ & 0.2120(5) & 0.3818(5) & 1/4 & 0.0429(19) \\
\hline
$x =$ 0.5 \\
Ba/Cs & 4$c$ & 1/2 & 0.1732(7) & 1/4 & 0.023(3)  \\
Fe & 8$e$ & 0.3540(5) & 1/2 & 0 & 0.0247(10) \\
Se1 & 4$c$ & 0 & 0.1195(5) & 1/4 & 0.0163(19) \\
Se2 & 8$g$ & 0.2153(5) & 0.3833(5) & 1/4 & 0.0467(19) \\
\hline
$x =$ 0.55 \\
Ba/Cs & 4$c$ & 1/2 & 0.1720(7) & 1/4 & 0.028(3) \\
Fe & 8$e$ &0.3539(4) & 1/2 & 0 & 0.0286(10) \\
Se1 & 4$c$ & 0 & 0.1209(5) & 1/4 & 0.0149(17) \\
Se2 & 8$g$ & 0.2177(5) & 0.3837(5) & 1/4 & 0.0522(19) \\
\hline
$x =$ 0.65 \\
Ba/Cs & 4$c$ & 1/2 & 0.1729(6) & 1/4 & 0.028(2) \\
Fe & 8$e$ & 0.3552(4) & 1/2 & 0 & 0.0214(8) \\
Se1 & 4$c$ & 0 & 0.1200(4) & 1/4 & 0.0175(15) \\
Se2 & 8$g$ & 0.2180(4) & 0.3840(4) & 1/4 & 0.0404(14) \\
\hline
$x = $ 0.75 \\
Ba/Cs & 4$c$ & 1/2 & 0.1714(7) & 1/4 & 0.031(2) \\
Fe & 8$e$ & 0.3563(3) & 1/2 & 0 & 0.0123(8) \\
Se1 & 4$c$ & 0 & 0.1205(4) & 1/4 & 0.0103(16) \\
Se2 & 8$g$ & 0.2208(4) & 0.3867(3) & 1/4 & 0.0244(13) \\
\end{tabular}
\end{ruledtabular}
\end{table}
\begin{table*}[htb]
\caption{\label{tab:table3}Refined crystallographic data taken at 300 K. Chemical formula units $Z$ is 4 for all $x$. Excluded regions contain peaks from carbon (2$\theta$ = 41.9-43.2$^\circ$ for $x = 0.65$ and 42.2-42.9$^\circ$ for $x = 0.75$), and unknown impurity (2$\theta$ = 24.5-25.3$^\circ$ and 47.8-48.7$^\circ$ for $x = 0.75$). $R_p = \sum_i \left| y_{{\rm obs}}-y_{{\rm calc}} \right| /\sum_i y_{{\rm obs}}$; $R_{wp} = \left[ \sum_i w_i \left| y_{{\rm obs}}-y_{{\rm calc}} \right| ^2/\sum_i w_i y_{{\rm obs}}^2 \right] ^{1/2}$.}
\begin{ruledtabular}
\begin{tabular}{cccccccc}
Formula & $x = 0.05$ & $x = 0.15$ & $x = 0.4$ & $x = 0.5$ & $x = 0.55$ & $x = 0.65$ & $x = 0.75$ \\
\hline
Molar mass (g/mol) & 485.7 & 485.2 & 484.1 & 483.7 & 483.5 & 483.0 & 482.6  \\
Space group & $Pnma$ & $Cmcm$ & $Cmcm$ & $Cmcm$ & $Cmcm$ & $Cmcm$ & $Cmcm$ \\
$a$ (\AA ) & 11.8648(3) & 9.2028(4) & 9.3242(3) & 9.3870(4) & 9.4083(3) & 9.4738(2) & 9.5885(2) \\
$b$ (\AA ) & 5.47187(9) & 11.7954(5) & 11.8173(5) & 11.8152(5) & 11.8342(4) & 11.8412(3) & 11.8475(3) \\
$c$ (\AA ) & 9.1775(2) & 5.5238(2) &  5.6091(2) & 5.6320(2) & 5.6413(1) & 5.6579(1) & 5.6785(1) \\
Cell volume (\AA$^3$) & 595.83(2) & 599.62(4) & 618.04(4) & 624.64(4) & 628.09(3) & 634.71(2) & 645.07(2) \\
$F(000)$ & 191.65 & 191.79 & 192.14 & 192.28 & 192.35 & 192.49 & 192.63 \\
Calculated & 5.415 & 5.375 & 5.203 & 5.144 & 5.113 & 5.055 & 4.969 \\
density (g/cm$^3$) \\
Number of & 30 & 23 & 23 & 23 & 23 & 23 & 23 \\%Refinement
parameters \\
Excluded regions & $-$ & $-$ & $-$ & $-$ & $-$ & 41.9-43.2 &  24.5-25.3;  \\
in 2 $\theta $ (deg.) &  &  &  &  &  &  & 42.2-42.9; 47.8-48.7 \\
No. of observed & 199 & 110 & 112 & 115 & 115 & 117 & 118  \\
reflections \\
$R_p (\%)$ & 14.2& 15.4 & 15.0 & 14.7 & 15.2 & 13.2 & 12.7 \\
$R_{wp} (\%)$ & 15.4 & 17.4 & 16.4 & 15.5 & 15.8 & 14.0 & 14.1 \\
$R_{\rm{exp}} (\%)$ & 3.72 & 3.78 & 4.13 & 4.42 & 3.89 & 3.96 & 4.53 \\
$\chi ^2$ &17.1 & 21.2 & 15.8 & 12.3 & 16.5 & 12.4 & 9.75 \\
\end{tabular}
\end{ruledtabular}
\end{table*}
A little amount of carbon contamination was found for $x = 0.65$ and $0.75$ samples, 0.55(1) \% and 0.027(1) \% weight ratio, respectively.
In addition, an unindexed impurity phase was detected for $x = 0.75$ sample, with the strongest peak which is roughly 20 times smaller than the main phase.
Although their peak intensities are very weak, they could affect the refinement results.
Hence, we excluded these 2$\theta$ regions: 41.9-43.2$^\circ$ (carbon) for $x = 0.65$, and 24.5-25.3$^\circ$, 47.8-48.7$^\circ$ (unknown impurity) and 42.2-42.9$^\circ$ (carbon)  for $x = 0.75$ during the refinements.

The space group of Ba$_{1-x} $Cs$_x$Fe$_2$Se$_3$ was determined using single crystal XRD.
We assumed that Ba$_{1-x} $Cs$_x$Fe$_2$Se$_3$ belong to either $Pnma$ or $Cmcm$, and that the structure boundary should exist between $x = 0$ and 1.
It should be noted that the presence of 112$_{Pnma}$ nuclear reflection, which is forbidden in the $Cmcm$, was used to determine the space group of Ba$_{1-x} $Cs$_x$Fe$_2$Se$_3$.
We found that the 112$_{Pnma}$ reflection was present for $x = 0.05$, but absent for $x \geq 0.15$ in the rotation photographs taken at the room temperature.
It is reported that BaFe$_2$Se$_3$ undergoes a phase transition from the $Pnma$ to $Cmcm$ structure at 660 K \cite{0953-8984-25-31-315403}.
Accordingly, it is natural to consider that the boundary of $Pnma$ and $Cmcm$ crosses 300 K between $x = 0.05$ and 0.15.

\begin{figure}[htb]
\includegraphics[width=1\hsize]{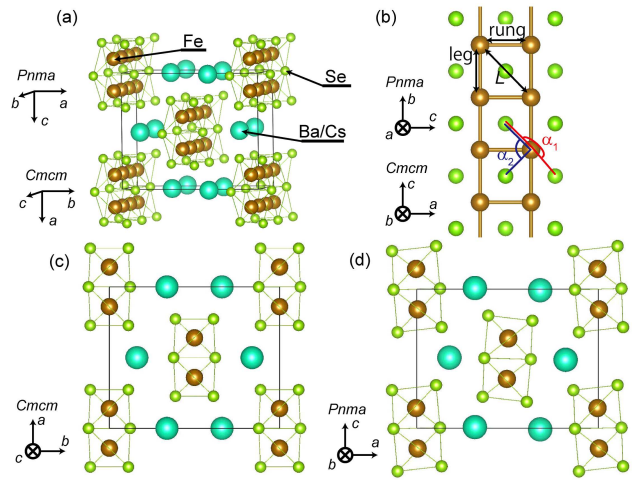}
\caption{(color online). (a), (b) Crystal structures of Ba$_{1-x}$Cs$_x$Fe$_2$Se$_3$. A cuboid with solid lines indicates a crystallographic unit cell. Note that the definition of $a, b$ and $ c$ axis is different from each other for the $Pnma$ and $Cmcm$ structures. Distinction for arrangement of ladder between (c) $Cmcm$ and (d) $Pnma$ space group. In $Pnma$, ladders are slightly tilted toward the $a$ axis.}
\label{atom_lattice_a-d}
\end{figure}

\begin{figure}[htb]
\includegraphics[width=1\hsize]{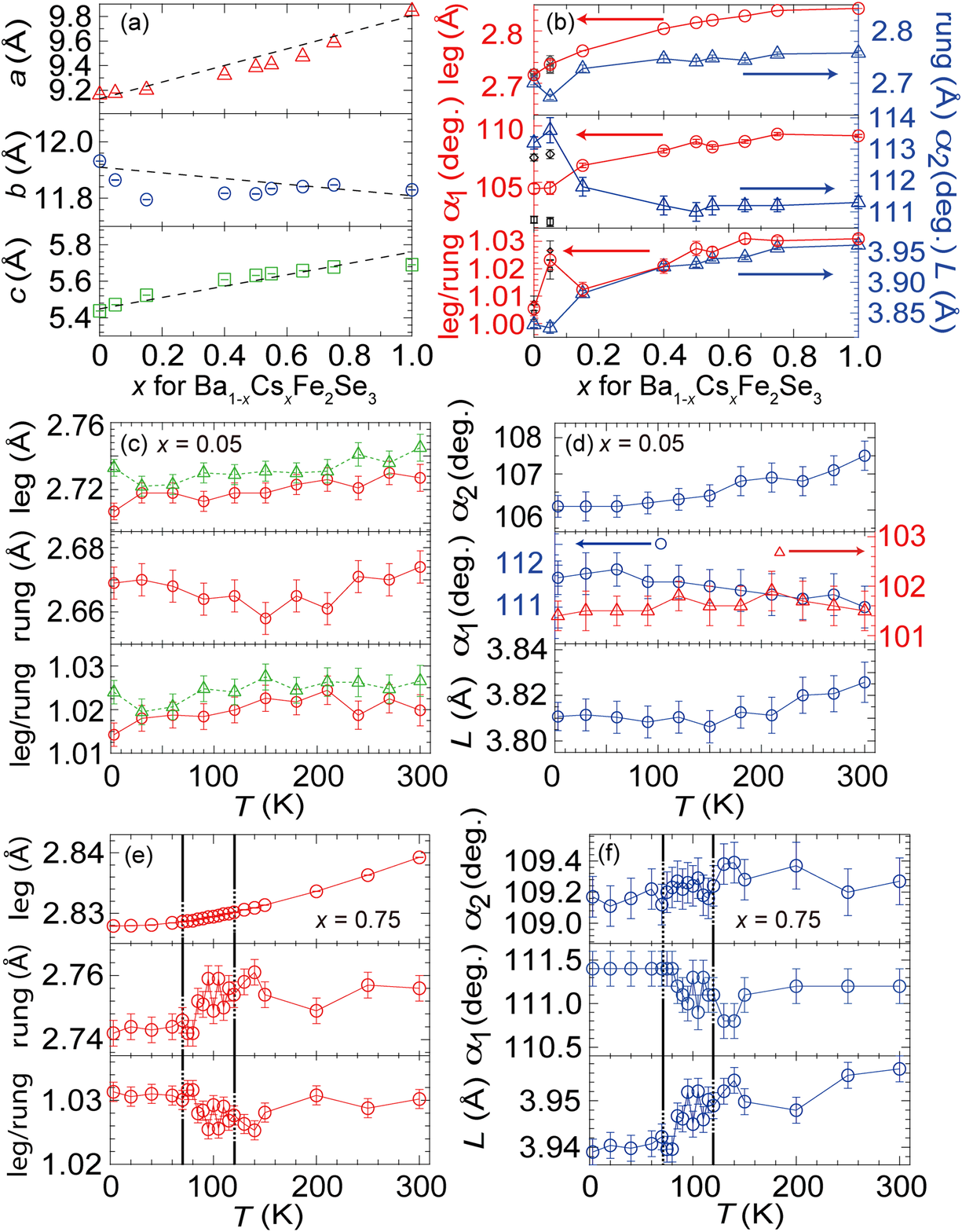}
\caption{(color online). (a) Cs concentration ($x$) dependence of lattice constants at 300 K. The data for $x = 0$ and $0.05$ are converted into the definition of $Cmcm$ symmetry. The dashed lines represent linear fit to the lattice constants. (b) The $x$ dependence of the Fe-Fe bond length along the leg and rung direction; Se-Fe-Se bond angle $\alpha _1$ and $\alpha _2$; the ratio between the bond length along the leg and rung direction, and the diagonal Fe-Fe bond length ($L$) [see Fig. \ref{atom_lattice_a-d}(b)]. (c), (d) $T$ dependence of local structures for $x$ = 0.05. Note that two distinct values for leg, leg/rung and $\alpha _1$ are owing to $Pnma$ space group (black square and rhomboid). The red circles show the average of the two values. (e), (f) $T$ dependence of local structures for $x$ = 0.75. Black lines indicate magnetic transition temperatures (see in text).}
\label{local_structure}
\end{figure}

Figure \ref{atom_lattice_a-d}(a) shows the crystal structure of Ba$_{1-x}$Cs$_x$Fe$_2$Se$_3$, and Fig. \ref{atom_lattice_a-d}(b) depicts one Fe ladder with both notation of $Pnma$ and $Cmcm$.
Figures \ref{atom_lattice_a-d}(c) and \ref{atom_lattice_a-d}(d) provide view being parallel to the leg direction of  $Cmcm$ and $Pnma$ structure, respectively.
The Fe ladders slightly incline in the $Pnma$ structure.
Since the definitions of $a, b$ and $c$ axes in $Pnma $ and $Cmcm$ are different, hereafter in this paper, the axes of Ba$_{1-x}$Cs$_x$Fe$_2$Se$_3$ are defined as follows: the $a$ axis is along the rung direction, the $b$ axis is perpendicular to the two-leg ladder plane, and the $c$ axis is along the leg direction, which is indeed the axis choice of the $Cmcm$ space group.
Figure \ref{local_structure}(a) shows $x$ dependence of lattice constants $a, b$ and $c$.
Among the three lattice constants, the $b$ value significantly deviates from the linearity in $0 < x < 0.5$.
In particular for $x = 0.15$ roughly 2 \% deviation was found.
It should be noted that this composition is close to the structural transition from the $Pnma$ to the $Cmcm$.
We also note that the decrease of the $b$ value indicates reduction of inter-ladder distance, giving rise to enhanced three dimensionality.
Figure \ref{local_structure}(b) shows that $x$ dependence of local structure parameters at 300 K.
The definition of the local structure parameters are given in Fig. \ref{atom_lattice_a-d}(b); the leg, rung, $L$ indicate the Fe-Fe bond lengths; $\alpha_1$ and $\alpha_2$ are the Se-Fe-Se bond angles. 
Local structures change continuously in $x \leq 0.15$ region, whereas weak anomaly could be seen around $x = 0.05$, which may be related to the structural phase transition.
Figures \ref{local_structure}(c), \ref{local_structure}(d) show temperature dependence of local structure parameters for $x =0.05$.
Although $x = 0.05$ is $Pnma $, being the same as BaFe$_2$Se$_3$, there is no enhancement of ladder distortion.
Figures \ref{local_structure}(e), \ref{local_structure}(f) show $T$ dependence of local structure parameters for $x = 0.75$.
Change in local structure was detected around $T = $ 120 and 70 K, where two successive magnetic phase transitions were observed from the paramagnetism to stripe-II, and from the stripe-II to stripe-I, respectively (described in \ref{neutron diffraction}).


\subsection{\label{resistivity_and_specific_heat}BULK PROPERTIES}
\begin{figure}[htb]
\includegraphics[width=1\linewidth]{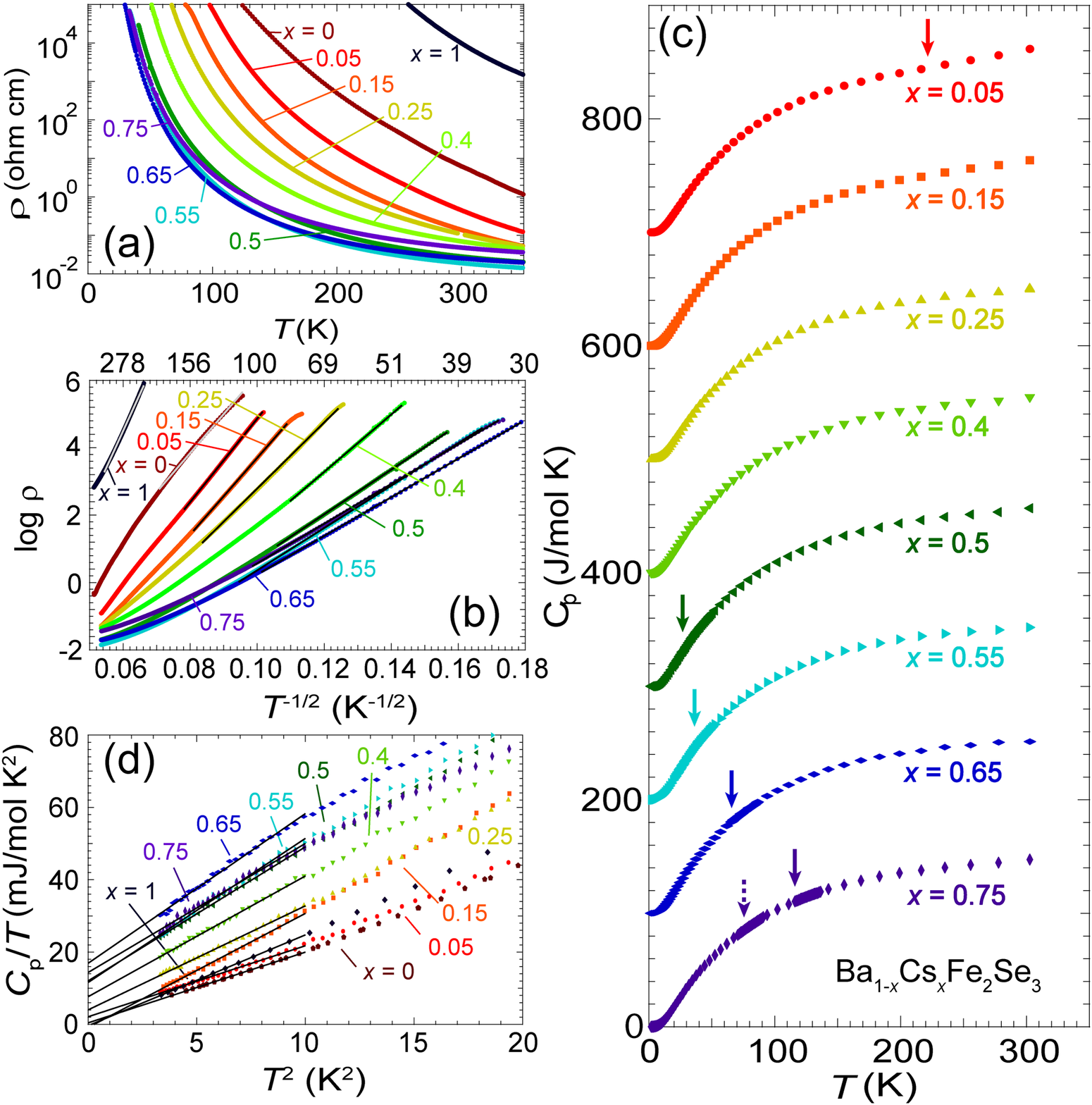}
\caption{(color online). (a) Temperature dependence of electrical resistivity of Ba$_{1-x} $Cs$_x$Fe$_2$Se$_3$ along the leg direction. (b) One-dimensional variable range hopping type plot of (a). Solid lines indicate the results of fitting. (c) Temperature dependence of the specific heat ($C_p$). Arrows show the magnetic transition temperature. Note that each data shifted by 100 $T/$mol K for clarity. (d) Specific heat divided by temperature against $T^2$ type plot. Solid lines indicate the results of fitting. Although Ba$_{1-x} $Cs$_x$Fe$_2$Se$_3$ show an insulating behavior, $\gamma$ term ($T$-linear term) remains.}
\label{phys_prop_latest_2}
\end{figure}
\begin{figure}[htb]
\includegraphics[width=1\linewidth]{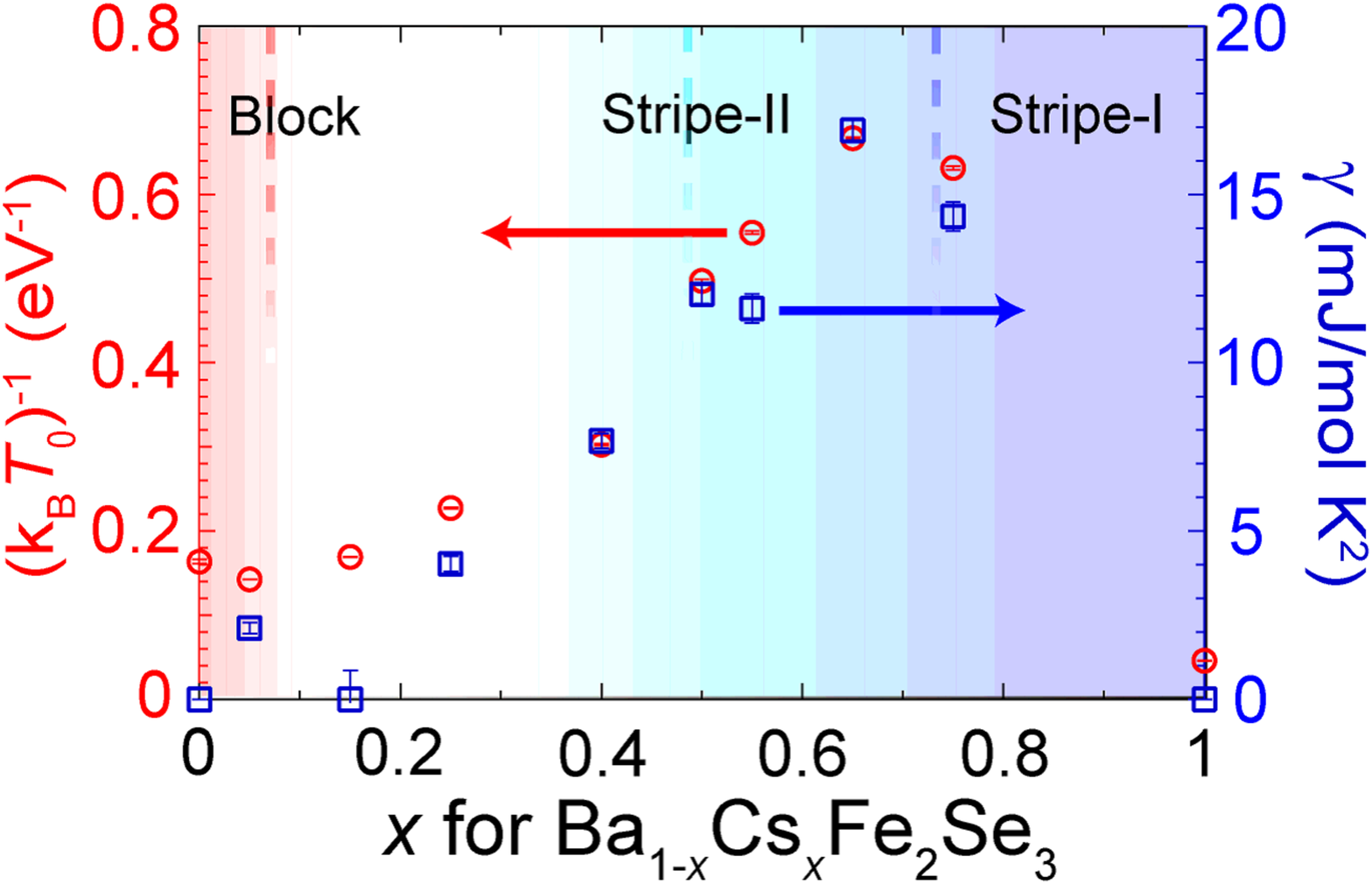}
\caption{(color online). Cs concentration ($x$) dependence of $\gamma$ (left-axis) and ($k_{\rm B}T_{\rm 0}$)$^{-1}$ (see text, right-axis). They show similar behavior and take the maximum value for $x$ = 0.65. In background, the type of magnetism at low temperature is shown.}
\label{gamma_T0_latest}
\end{figure}
\begin{figure}[htb]
\includegraphics[width=0.75\linewidth]{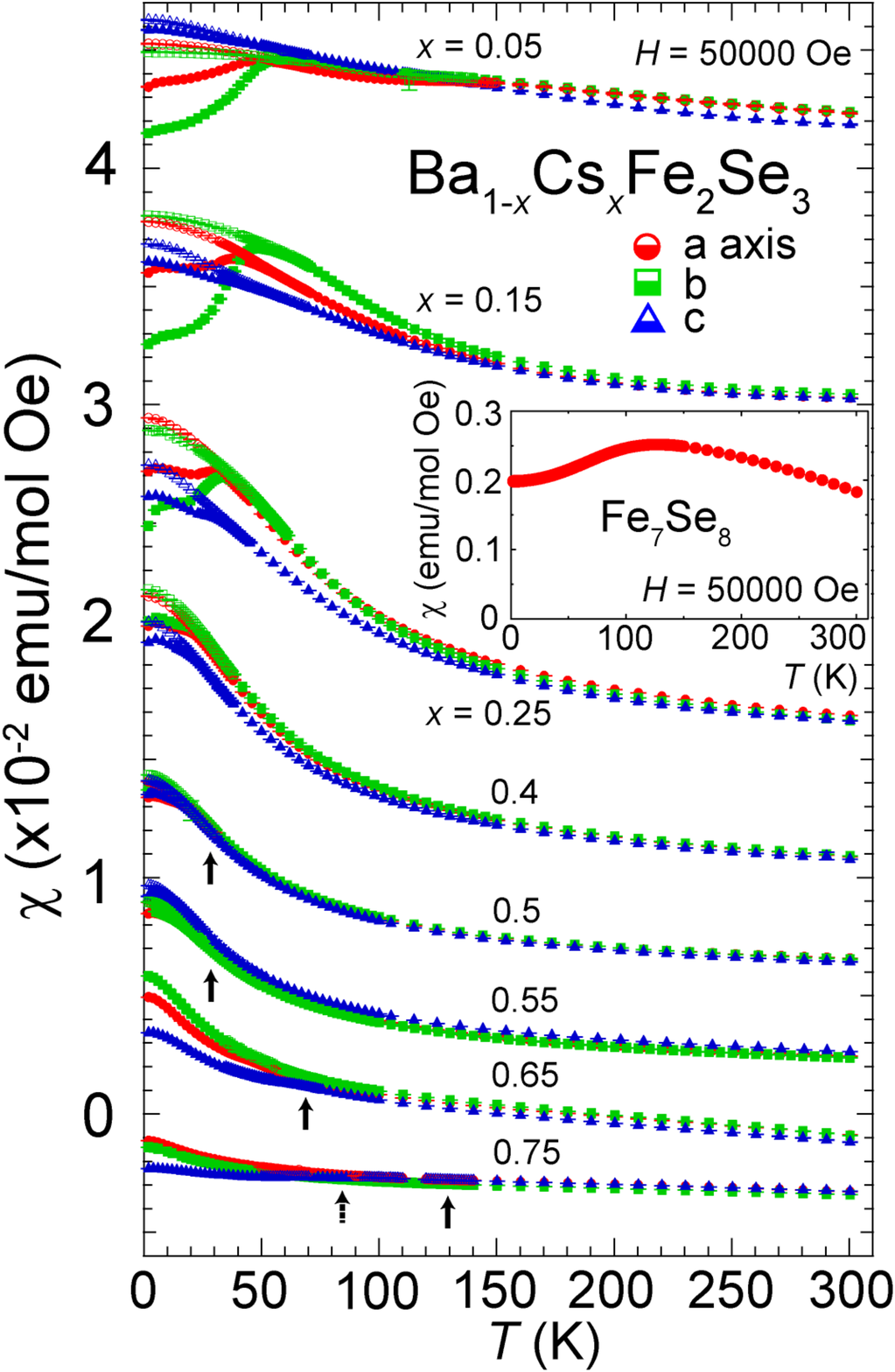}
\caption{(color online). Temperature ($T$) dependence of magnetic susceptibility ($\chi$) for Ba$_{1-x} $Cs$_x$Fe$_2$Se$_3$ along the $a, b$ and $c$ axis in the $Cmcm$ notation at the magnetic field $H$ = 50000 Oe. The solid and open symbols indicate zero-field cooled (ZFC) and field cooled (FC) measurements, respectively. Note that each data are shifted for clarity. The arrows indicate the magnetic transition temperatures. The inset shows temperature dependence of $\chi$ for the magnetic impurity Fe$_7$Se$_8$.}
\label{chi_M_sum}
\end{figure}

Temperature dependence of electrical resistivity ($\rho $) along the leg direction is shown in Fig. \ref{phys_prop_latest_2}(a).
No anomaly was seen in $\rho $ in the whole measured temperature range.
Intermediate compounds show much lower $\rho$ than that of the parent compounds; at $x = 0.55$, $\rho$ at 300 K shows the lowest value $\rho = 1.89 \times 10^{-2} $ $\Omega$ cm, which is 10$^2$ times smaller than that of BaFe$_2$Se$_3$ and 10$^5$ smaller than that of CsFe$_2$Se$_3$.
The temperature dependence of $\rho $ shows one-dimensional variable-range-hopping (VRH) behavior in low-$T$ region, given by $\rho \propto \exp [(T_0 / T)^{1/2}] $ [Fig. \ref{phys_prop_latest_2}(b)] \cite{NFMott, doi:10.1080/14786437208226975, BIShklovskii}.
It is known that $(k_{\rm{B}} T_0)^{-1}$ of 1D VRH is proportional to the localization length and the density of states at Fermi level.
Figure \ref{gamma_T0_latest} shows the $(k_{\rm{B}} T_0)^{-1}$ obtained by fitting the data for Ba$_{1-x} $Cs$_x$Fe$_2$Se$_3$ versus $x$.
Although $\rho $ of CsFe$_2$Se$_3$ ($ x = 1$) \cite{PhysRevB.85.214436} are deviated from the 1D VRH behavior, we fit it with 1D VRH to compare $(k_{\rm{B}} T_0)^{-1}$ with those of Ba$_{1-x} $Cs$_x$Fe$_2$Se$_3$.
Interestingly, the largest $(k_{\rm{B}} T_0)^{-1}$ was observed for $x = 0.65$, which is closer to the strong insulator CsFe$_2$Se$_3$ than BaFe$_2$Se$_3$.

Figure \ref{phys_prop_latest_2}(c) shows specific heat ($C_p$) as a function of temperature.
No anomaly was observed in all measured temperature and all $x$ ranges.
The $C_p/T$ versus $T^2$ plots at low temperature indicate $T$-linear contribution ($\gamma$ term) extrapolated as $T \to 0$ except for $x = 0.15$ [Fig. \ref{phys_prop_latest_2}(d)].
This suggests that Ba$_{1-x} $Cs$_x$Fe$_2$Se$_3$ have finite $\gamma$ values.
As discussed below, magnetization measurements indicate that the samples contain Fe$_7$Se$_8$.
One can suspect that the obtained specific heat contain the contribution from the impurity.
However, the amount of Fe$_7$Se$_8$ is at most 1 \% molar ratio (see below) and its $\gamma$ values are relatively small; 8 and 4.4 mJ$/$mol K$^2$ for 4$c$ and 3$c$ structures of Fe$_7$Se$_8$, respectively \cite{doi:10.1143/JPSJ.16.1162, PhysRevB.57.8845}.
Therefore the impurity effect is negligible.
The $\gamma$ values obtained by linear fit for $T < 3.3$ K [solid lines in Fig. \ref{phys_prop_latest_2}(d)] are summarized in Fig. \ref{gamma_T0_latest}.
We note that ($k_{\rm B}T_{\rm 0}$)$^{-1}$ and $\gamma$ show a similar $x$ dependence.

Figure \ref{chi_M_sum} shows the magnetic susceptibilities ($\chi$) defined as magnetization ($M$) over magnetic field ($H$), that is, $\chi \equiv M/H$, for single crystals of Ba$_{1-x} $Cs$_x$Fe$_2$Se$_3$ under $H $ = 50000 Oe.
The samples contain a tiny amount of ferromagnetic impurity, and in this case, Fe$_7$Se$_8$ is the most plausible one.
Estimated molar ratio of Fe$_7$Se$_8$ in Ba$_{1-x} $Cs$_x$Fe$_2$Se$_3$ from hysteresis loop is less than 1 \% for $x = 0.05$, and 0.5 \% for other compositions \cite{magnetization_impurity}.
We also measured the temperature dependence of magnetization for the annealed Fe$_7$Se$_8$ between $T$ = 2 and 300 K on field cool (FC) and zero-field cool (ZFC) measurements [inset of Fig. \ref{chi_M_sum}], and found that $\chi$ decreases below 120 K on cooling.
However $\chi$ of Ba$_{1-x} $Cs$_x$Fe$_2$Se$_3$ do not show such behavior, indicating that the negligible contribution from Fe$_7$Se$_8$ impurity phase.
The $\chi$ of Ba$_{1-x} $Cs$_x$Fe$_2$Se$_3$ show paramagnetic increase as temperature decreases, and glassy behaviors were observed for $0.05 \leq x \leq 0.4$ at low temperatures.
The systematic decrease of the spin-glass transition temperature ($T_f$) implies that this behavior is of intrinsic nature; if this behavior were due to ferromagnetic impurity phase, $T_f$ would be constant.

\subsection{\label{neutron diffraction}MAGNETIC STRUCTURES}
To elucidate variation of magnetic structures in Ba$_{1-x} $Cs$_x$Fe$_2$Se$_3$, we performed powder neutron diffraction measurements.
Figures \ref{result_1}, \ref{result_3} and \ref{result_2} show the powder diffraction patterns of several temperatures from 3 to 300 K, in a selected scattering wave vector $Q$ range.
For $x = 0.05, 0.4, 0.65$ and $0.75$ the diffraction patterns were collected using ECHIDNA, and for $x = 0.15, 0.25, 0.5$ and $0.55$ using WAND.
\subsubsection{$0 < x \leq 0.15$}
\begin{figure}[htb]
\includegraphics[width=1.1\linewidth]{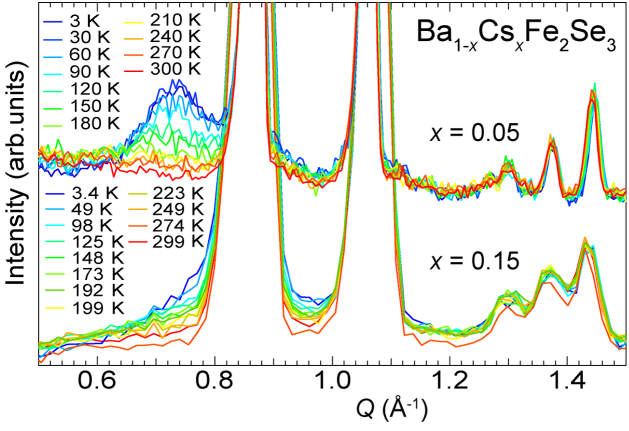}
\caption{(color online). Powder neutron diffraction patterns for $x = 0.05$ and 0.15 taken at several temperature points. The data were collected at the listed temperatures using ECHIDNA for $x = 0.05$ and WAND for $x = 0.15$.}
\label{result_1}
\end{figure}
\begin{figure}[htb]
\includegraphics[width=1.1\linewidth]{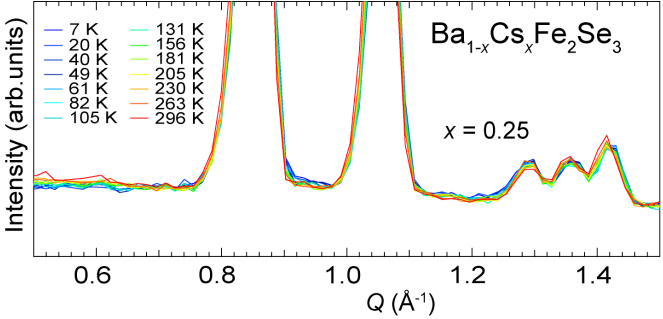}
\caption{(color online). Powder neutron diffraction patterns for $x = 0.25$  taken at several temperature points. The data were collected at the listed temperatures using WAND.}
\label{result_3}
\end{figure}
\begin{figure}[htb]
\includegraphics[width=1\linewidth]{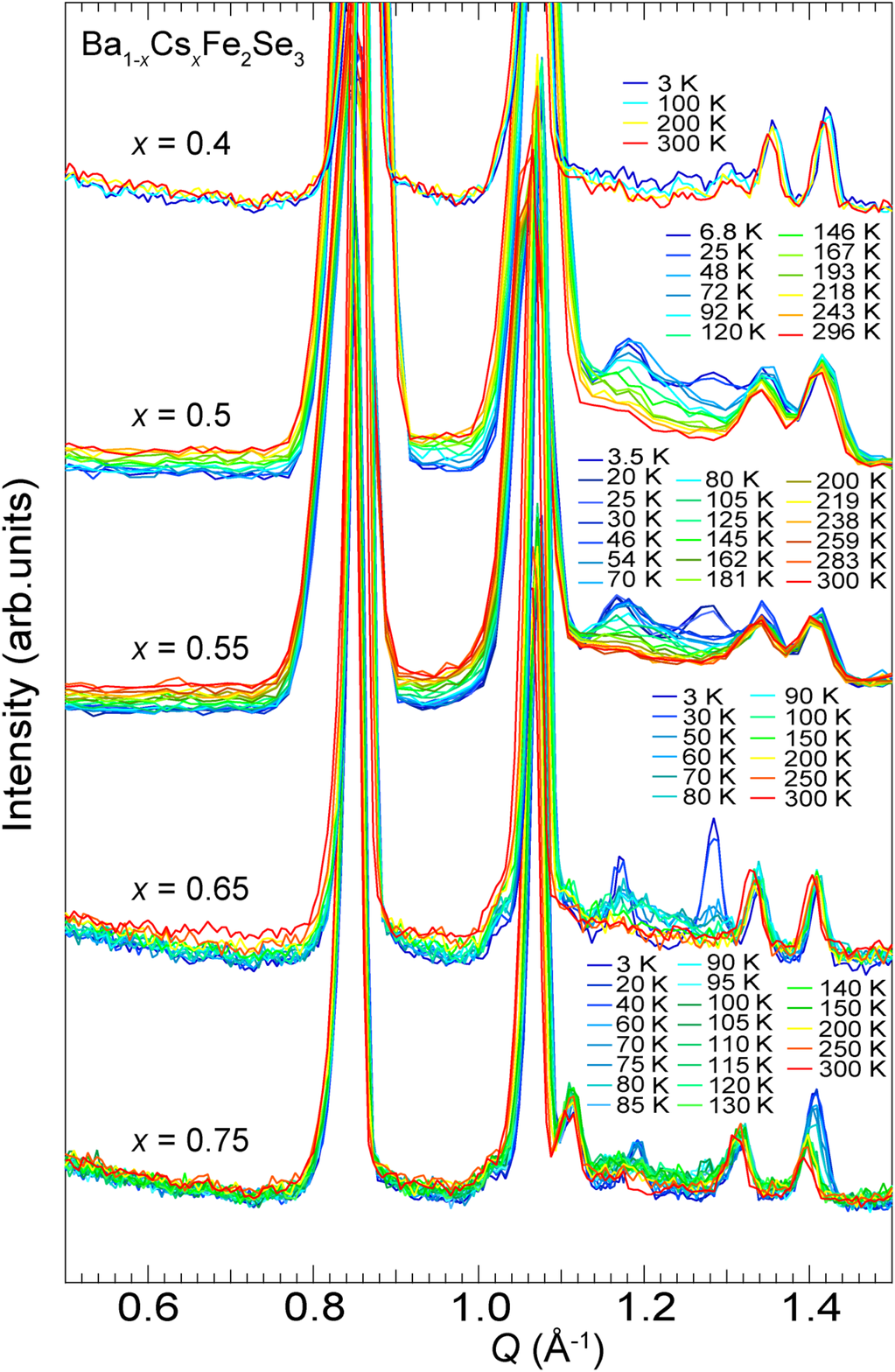}
\caption{(color online). Powder neutron diffraction patterns for $x = 0.4$  taken at several temperature points, 0.5, 0.55, 0.65, and 0.75. The data were collected at the listed temperatures using ECHIDNA for $x = 0.4$, 0.65 and 0.75, and using WAND for $x = 0.5$ and 0.55.}
\label{result_2}
\end{figure}
The diffraction patterns for $x = 0.05$ and 0.15 are shown in Fig. \ref{result_1}.
For $x = 0.05$ magnetic reflection appears at $Q \sim 0.7$ \AA$^{-1} $, which corresponds to the main peak of block magnetism in BaFe$_2$Se$_3$ appearing at the magnetic wave vector \mbox{\boldmath $q$}$_m$ = (0.5, 0.5, 0.5) \cite{PhysRevB.85.064413}.
This magnetic scattering appears below $T$* = 290(10) K, and grows with decreasing temperature.
This peak is broadened beyond instrumental resolution even at the lowest temperature, suggesting that only 5 \% doping eliminates the long ranged correlation of the block magnetism.
Likewise, for $x = 0.15$, only magnetic diffuse scattering was observed below $T$* = 215(15) K at similar position to the block magnetism.
The correlation length is roughly estimated from the peak width as: 16(4) \AA \  for $x = 0.05$ at 3 K, 9(3) \AA \  for $x = 0.15$ at 3.4 K.
BaFe$_2$Se$_3$ ($x = 0 $) have 19(5) \AA \  correlation length at 275 K \cite{PhysRevB.85.064413}, thus these correlation lengths seem to decrease toward $x = 0.25$, where the magnetic orders are completely suppressed.

\subsubsection{$0.4 \leq x < 1$}

\begin{figure}[htb]
\includegraphics[width=1\hsize]{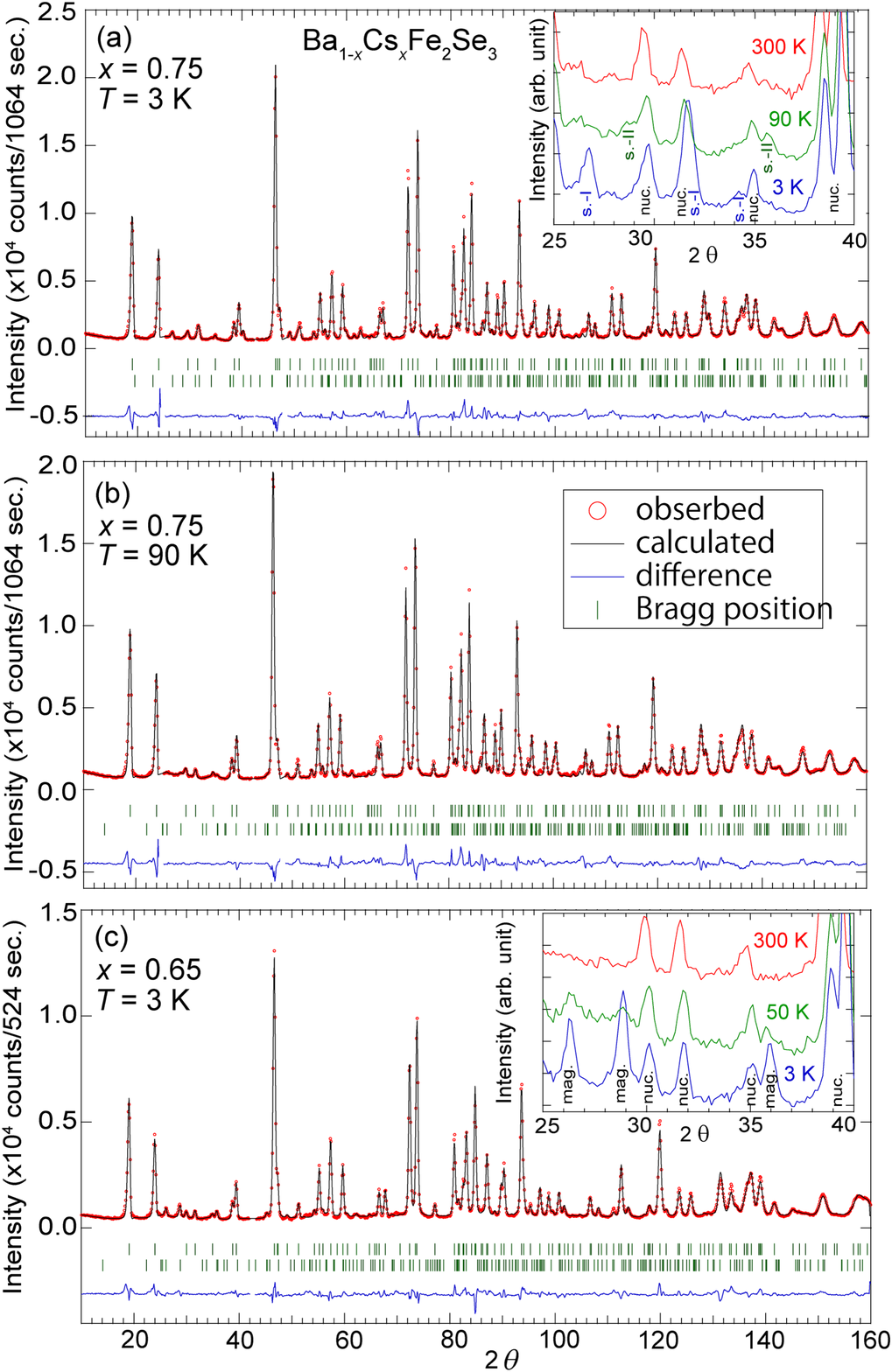}
\caption{(color online). High-resolution powder neutron diffraction patterns for (a) $x = 0.75$ at 3 K, (b) $x = 0.75$ at 90 K,
and (c) $x = 0.65$ at 3 K using ECHIDNA with Rietveld analysis results (solid lines). The calculated positions of nuclear and magnetic reflections are indicated (green ticks). The bottom line gives the difference. The insets show (a) change of positions of magnetic reflections from stripe-II to stripe-I magnetism and (c)  temperature evolution of magnetic reflections for $x = 0.65$.}
\label{rietveld_sum}
\end{figure}

Figure \ref{result_2} shows diffraction patterns for $0.4 \leq x < 1$, showing stripe magnetism.
We first focus on $x = 0.75$.
On cooling, a magnetic diffuse scattering develops below 250 K at $Q = 1.2$ \AA$^{-1} $.
In contrast to the block magnetism, magnetic reflections become almost resolution limited below 130(5) K.
These magnetic reflections change their positions below 80(5) K, which corresponds to additional magnetic transition [inset in Fig. \ref{rietveld_sum}(a)].

\begin{figure}[htb]
\includegraphics[width=1\linewidth]{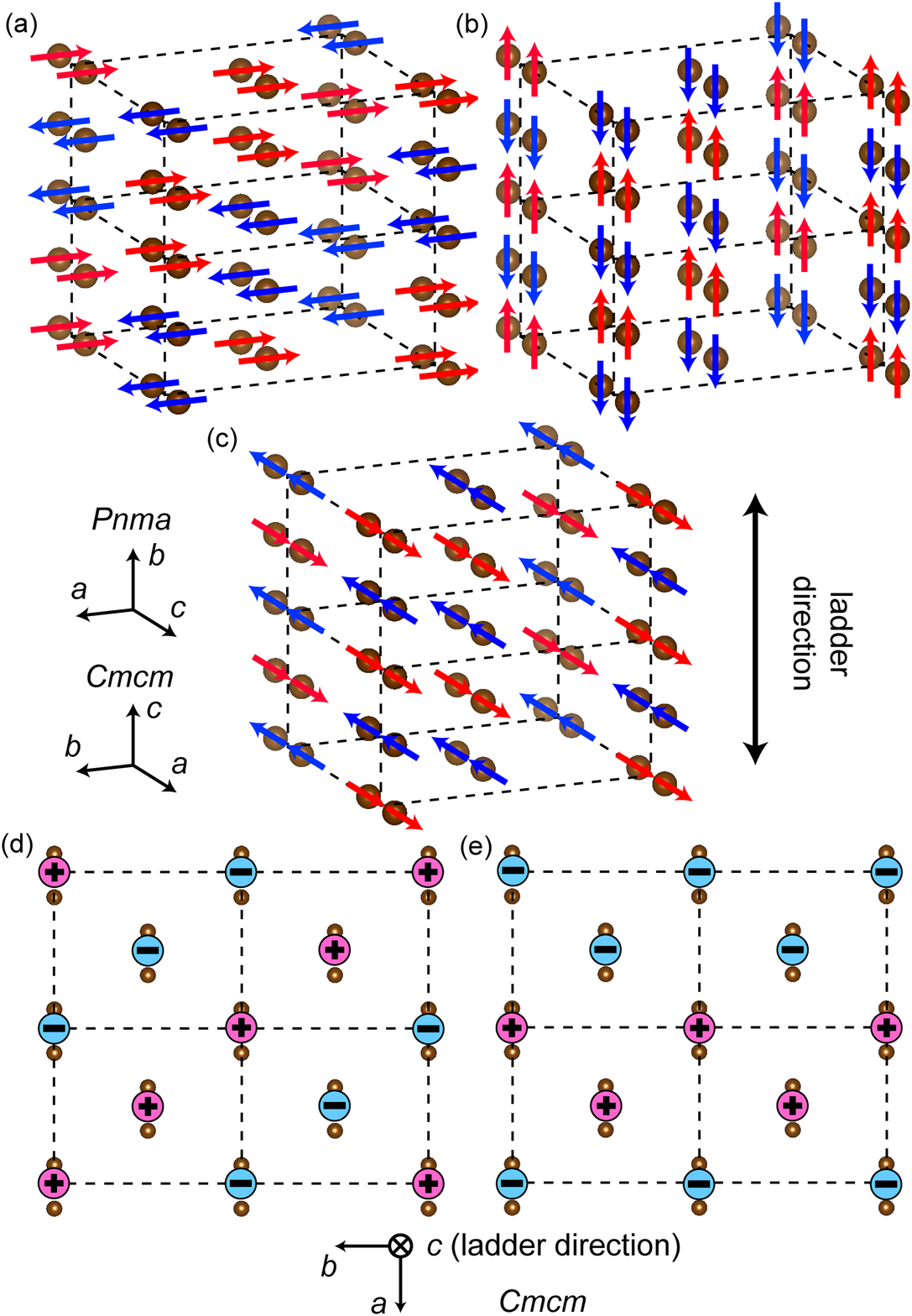}
\caption{(color online). The schematic figures of magnetic structures for: (a) block magnetism (b) stripe-I magnetism, and (c) stripe-II magnetism. The inter-ladder relation of spins are shown in (d) stripe-I and (e) stripe-II magnetism. The $+$ and $-$ sign indicate ferromagnetic and antiferromagnetic correlation in neighboring ladders, respectively.}
\label{mag_structure_final}
\end{figure}

All the magnetic peak positions below 80(5) K are well accounted for by the propagation vector \mbox{\boldmath $q$}$_m$ = (0.5, 0.5, 0), which is the same as CsFe$_2$Se$_3$.
To identify this magnetic structure, we applied representation analysis and performed Rietveld analysis.
Basis vectors (BVs) of the irreducible representations (irreps) of \mbox{\boldmath $q$}$_m$ were calculated using the SARA$h$ code \cite{Wills2000680}.
The obtained BVs are listed in Table II in Ref. \cite{PhysRevB.85.214436}.
We sorted all BVs by comparing $R$-factors and found that $\psi_9 $, with spins aligned along the leg direction, has the best fit with  $R_p = 11.7$ \%; the second best is 12.8 $ \% $ for $\psi_2 $.
The Rietveld refinement results are shown in Fig. \ref{rietveld_sum}(a).
The obtained magnetic structure is shown in Fig. \ref{mag_structure_final}(b).
We call this magnetic structure as stripe-I magnetism.
This magnetic structure is completely the same as that of CsFe$_2$Se$_3$, implying the relatively stable stripe-I magnetism against the Ba substitution.

On the other hand, between 130(5) and 80(5) K magnetic reflection positions cannot be accounted for by the propagation vector of BaFe$_2$Se$_3$, \mbox{\boldmath $q$}$_m$ = (0.5, 0.5, 0.5) nor of CsFe$_2$Se$_3$, \mbox{\boldmath $q$}$_m$ = (0.5, 0.5, 0).
The peak positions are instead well accounted for by \mbox{\boldmath $q$}$_m$ = (0.5, 0, 0), however peak intensities of the magnetic reflections are too weak to perform Rietveld refinements.
With decreasing $x$, we found that the new magnetic phase becomes the ground states in $x$ = 0.5, 0.55 and 0.65 above 3 K and that the reflections are the most intense for $x$ = 0.65.
Therefore we analyzed the magnetic structure for $x = 0.65$.
Figure \ref{rietveld_sum}(c) shows a diffraction pattern for  $x = 0.65$ taken at 3 K and a result of Rietveld refinement.
We performed representation analysis in the same way as $x = 0.75$ and obtained BVs summarized in Table \ref{x=065_BVs}.
The sites of Fe atom in orbit 1 and orbit 2 cannot be exchanged by symmetry operations that leave the \mbox{\boldmath $q$}$_m$ invariant.
Therefore, we can choose different phase factors of moments for Fe atoms in orbit 1 and orbit 2.
Magnetic moment of the $j$-th Fe atom \mbox{\boldmath $m$}$_{j}$  ($j = 1, 2 \in \mbox{orbit1}$ or $j = 3, 4 \in \mbox{orbit2}$) is given by the real part of $\mbox{\boldmath $m$}_{j \in \mbox{orbit1}} = C \psi_1^{j \in \mbox{orbit1}} e^{\phi_1 i} e^{2\pi i \mbox{\boldmath $q$}_m \cdot \mbox{\boldmath $t$} } $ or $\mbox{\boldmath $m$}_{j \in \mbox{orbit2}} = C \psi_1^{j \in \mbox{orbit2}} e^{\phi_2 i} e^{2\pi i \mbox{\boldmath $q$}_m \cdot \mbox{\boldmath $t$} } $
, where $\psi_1^{\in \mbox{orbit $i$}}$ means BV $\psi_1$ for $j$-th Fe atom belongs to orbit $i$ ($i = 1, 2$), $\phi_i$ is phase factor of orbit $i$, \mbox{\boldmath $t$} is the lattice translation vector and $C$ is a real constant.
We assumed that orbit 1 and orbit 2 are in the same irreps and that all Fe atoms have the same moment size, because Fe atoms in paramagnetic state are in single site (see Table \ref{tab:table2}).
The best fit is a combination of $\psi_1$ with the phase factor $\phi_1 = 5\pi / 4$ and $\phi_2 = \pi/4 $, where $R_p = 12.0$ \% (the second best is 12.2 $\%$ for $\psi_5$).

The schematic magnetic structure is shown in Fig. \ref{mag_moment_2}(c).
We call this magnetic structure as stripe-II.
Stripe-I and stripe-II have similar intra-ladder structure, where magnetic moments are arranged to form ferromagnetic units along the rung direction, and the units stack antiferromagnetically along the leg direction.
In contrast, the spin directions [Fig. \ref{mag_structure_final}(b) and \ref{mag_structure_final}(c)] and the inter-ladder relation [Fig. \ref{mag_structure_final}(d) and \ref{mag_structure_final}(e)] are different between stripe-I and stripe-II.

\begin{table}[htb]
\caption{Basis vectors (BVs) of irreducible representations (irreps) for the space group $Cmcm$ with the magnetic wave vector \mbox{\boldmath $q$}$_m$ = (0.5, 0, 0). Subscripts show the moment direction. Columns for positions represent No.1: $(x, 1/2, 0)$, No. 2: $(x, 1/2, 1/2)$, No. 3: $(1-x, 1/2, 1/2)$ and No. 4: $(1-x, 1/2, 0)$.}
\label{x=065_BVs}
\begin{ruledtabular}
\begin{tabular}{ccccccc}
Irreps & BV & \multicolumn{2}{c}{Orbit 1} & \multicolumn{2}{c}{Orbit 2} \\
& & No. 1 & No. 2 & No. 3 & No. 4 &  \\
\hline
$\Gamma_1$ & $\psi_1$ & 2$_a$ & -2$_a$ & 2$_a$ & -2$_a$ \\
$\Gamma_2$ & $\psi_2$ & 2$_a$ & 2$_a$ & 2$_a$ & 2$_a$ \\
$\Gamma_3$ & $\psi_3$ & 2$_b$ & 2$_b$ & 2$_b$ & 2$_b$ \\
 & $\psi_4$ & 2$_c$ & -2$_c$ & 2$_c$ & -2$_c$ \\
$\Gamma_4$ & $\psi_5$ & 2$_b$ & -2$_b$ & 2$_b$ & -2$_b$ \\
 & $\psi_6$ & 2$_c$ & 2$_c$ & 2$_c$ & 2$_c$ \\
\end{tabular}
\end{ruledtabular}
\end{table}

Knowing the magnetic structure of the $x = 0.65$ sample, we tried to determine the magnetic structure for $x = 0.75$ at 90 K, where the same initial magnetic structure is assumed.
The best fits are obtained for $\psi_4$ and $\psi_5$ with $R_p = 11.7$ \%, however $\psi_1$ have the close value with  $R_p = 11.8$ \%.
The $\chi$ [Fig. \ref{chi_M_sum}(b)] show little anisotropy for $ 0.5 \leq x \leq 0.75$.
Nevertheless it is hard to expect any anisotropic change without variations in significant structural parameter for $x = 0.65 $ and $x = 0.75$, we speculate that the spin orientation is parallel ($\psi_1$)  also at 90 K of the $x = 0.75$ sample.
To confirm this orientation, single crystal neutron diffraction measurements is desired.


For $0.4 \leq x \leq 0.75$, magnetic diffuse scatterings were also observed at $Q \sim 1.2 $ \AA$^{-1}$ at higher temperatures.
These scatterings evolve below $\sim 200 $K [see rhomboids in Fig. \ref{phase_diagram_latest_2}].
The positions are different from that of the $x = 0.05$ and 0.15 samples, where diffuse scatterings appear at $ Q \sim 0.7$ \AA$^{-1}$ near the Bragg reflection position of the block magnetism.
Therefore these diffuse scatterings would reflect stripe-type short-range correlation.
No diffuse scattering was observed in CsFe$_2$Se$_3$.

\subsubsection{Magnetism of $x = 0.25$}

Interestingly, for $x = 0.25$, no magnetic reflection was observed down to 7 K as shown in Fig. \ref{result_3}.
The $\chi$ for $x = 0.25$ shows the spin-glass behavior, and it is consistent with this result.
Note that other intermediate compounds ($0.05 \leq x \leq 0.4 $) also show spin-glass behavior, however they have some magnetic scatterings (mostly diffuse) in powder neutron diffraction. 

\subsubsection{Moment size of Ba$_{1-x} $Cs$_x$Fe$_2$Se$_3$}

\begin{figure}[htb]
\includegraphics[width=1\linewidth]{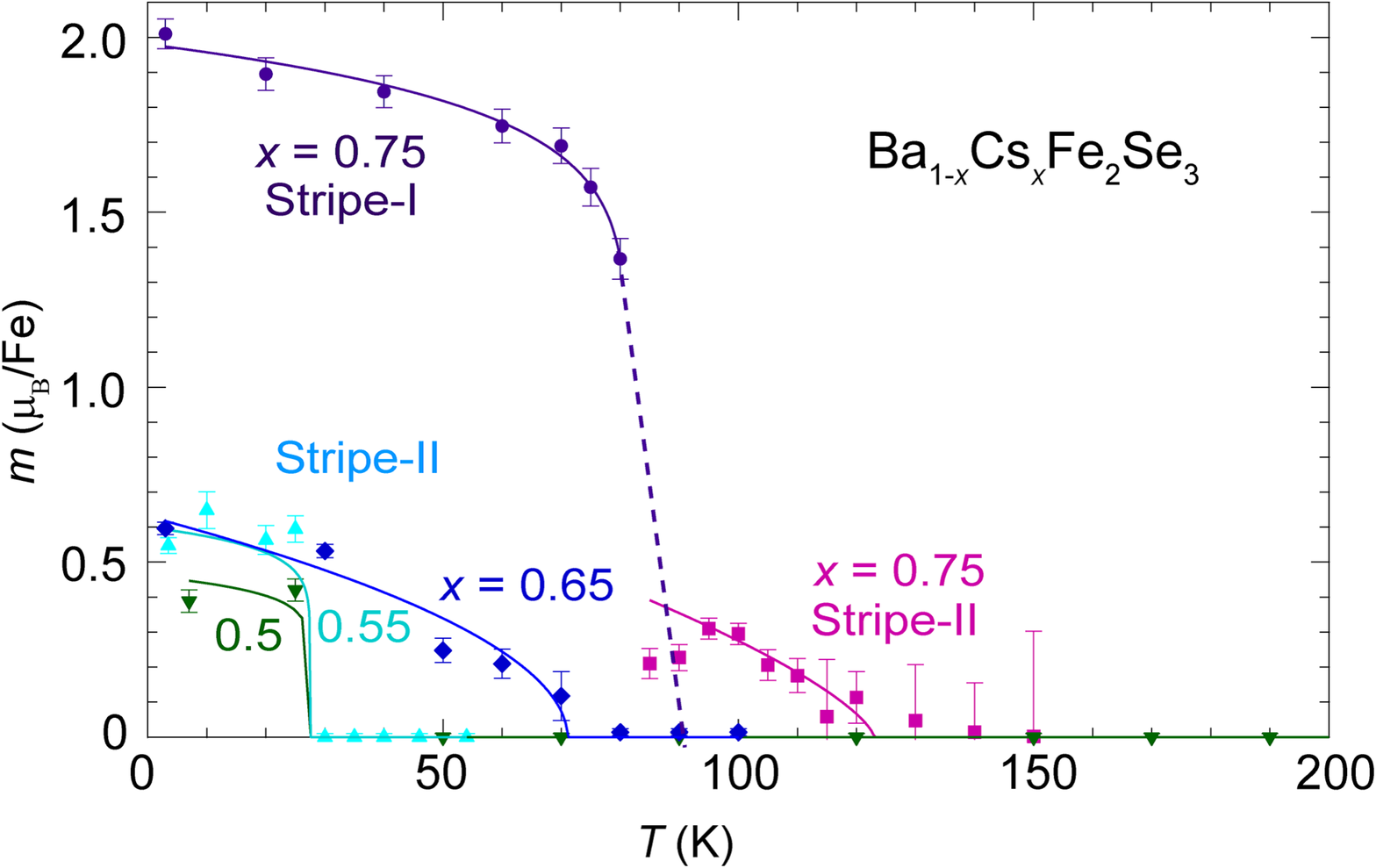}
\caption{(color online). Temperature dependence of estimated magnetic moment. The data of $x = $0.65 and 0.75 were collected using ECHIDNA, and of $x = $ 0.5 and 0.55 using WAND.}
\label{mag_moment_2}
\end{figure}

Temperature dependence of estimated magnetic moment sizes for several $x$ are shown in Fig. \ref{mag_moment_2}.
The moment sizes of $x =0.65$ and 0.75 are the results of Rietveld refinement.
For $x = 0.5$ and 0.55 moment sizes were obtained from comparing between integrated intensity of the 021$_{Cmcm}$ nuclear peak and the $\frac{1}{2}$21$_{Cmcm}$ magnetic peak.
The $x = 0.75$ sample shows successive phase transitions.
Around 130 K the stripe-II magnetism appears and the magnetic moment grows to 0.31(3) $\mu_{\rm B}$.
In 85 - 95 K, the stripe-I and stripe-II structures could coexist, because in this temperature range the moment size of stripe-II becomes small.
On further cooling, the stripe-I magnetism is finally stabilized, and the magnetic moment reaches to 2.01(5) $\mu_{\rm B}$ at 3 K.
The $x = 0.5, 0.55$ and 0.65 samples, where stripe-II magnetism appears, show similar moment size to the $x = 0.75$ sample at the lowest temperature; 0.39(3) $\mu_{\rm B}$ for $x = 0.5$; 0.55(3) $\mu_{\rm B}$ for $x = 0.55$; and 0.60(2) $\mu_{\rm B}$ for $x = 0.65$.
As $x$ decreases to the region where magnetism is suppressed, the moment size decreases.

In iron-based SCs, magnetic structure shows a close correlation with the ordered moment.
Single-stripe magnetism usually has an ordered moment smaller than 1 $\mu_{{\rm B}}$, double-stripe has $\sim$ 2 $\mu_{{\rm B}}$ and block magnetism has a larger moment around 3 $\mu_{{\rm B}}$.
On the other hand for iron-based ladder compounds, stripe-II magnetism has $\sim $ 0.5 $\mu_{{\rm B}}$, stripe-I has $ \sim$ 2 $\mu_{{\rm B}}$ and block magnetism has $ \lesssim$ 3 $\mu_{{\rm B}}$.
The correlations are basically consistent between the 2D iron-based SCs and the ladder compounds.
It may be noteworthy that the ladder compounds cover whole range of the moment size seen in iron-based SCs.

\subsubsection{Magnetic phase diagram}
\begin{figure}[htb]
\includegraphics[width=1\linewidth]{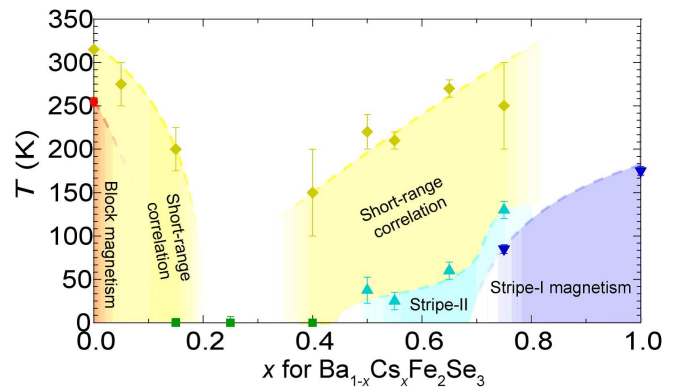}
\caption{(color online). Magnetic phase diagram determined by the powder neutron diffraction. 
The data were collected down to 3 K ($x = 0.05, 0.4, 0.65$, and 0.75), 5K ($x = 0.15, 0.5$ and 0.55), and 7 K ($x = 0.25$). Rhomboids indicate the temperature where diffuse scattering appears. Squares, upward triangles, and downward triangles indicate the magnetic phase transition temperature of the block, stripe-I, and stripe-II magnetism, respectively.}
\label{phase_diagram_latest_2}
\end{figure}

Figure \ref{phase_diagram_latest_2} shows established temperature and composition phase diagram of Ba$_{1-x} $Cs$_x$Fe$_2$Se$_3$.
Interestingly, $T_{\rm{N}} $ is drastically suppressed by increasing $x$, and only 5 \% Cs doping ($i.e.$ $x = 0.05$) destabilizes the long range order of the block magnetism.
Almost simultaneously, space group changes from $Pnma$ to the $Cmcm$ between $x = 0.05$ and 0.15.
Compared with the block magnetism, stripe-I and II magnetism can be seen in a large region of phase diagram and are relatively resilient the Cs doping.
For $x = 0.75$, successive magnetic phase transition from the paramagnetic state to stripe-II and from stripe-II to stripe-I was observed.
With decreasing Cs concentration $x$, the stripe-II magnetism becomes ground state for $0.5 \leq x \leq 0.65$ and $T_{\rm N}$ decreases gradually.
The long range magnetic order is completely suppressed at $x = 0.4$.

\section{\label{discussion}DISCUSSION}

We here discuss the interplay of crystal structure, bulk properties, and magnetism of Ba$_{1-x} $Cs$_x$Fe$_2$Se$_3$.

The $T$-linear contribution ($\gamma$ value) of $C_p$ becomes large around $x = 0.65$ (Fig. \ref{gamma_T0_latest}).
Possible origin of finite $\gamma$ values in $C_p$ may be spin-glass state, magnons or electrons.

Firstly, for the spin-glass possibility, we obtained the highest $T_f$ for $x = 0.15$, and for $0.5 \leq x \leq 0.75$ glassy behavior did not appear [Fig. \ref{chi_M_sum}(b)].
This is not consistent with the behavior of $\gamma$ values, thus the possibility of spin-glass contribution can be excluded.

Secondly, temperature dependence of $C_p$ is known as $C_p \propto T^{d/\nu} $ for magnetic excitations associated with magnons, where $d$ is spatial dimensionality, $\nu =$ 1 for antiferromagnetic; $= 2$ for ferromagnetic ordered system. 
Thus $T$-linear contribution of the magnons could be due to 1D antiferro-type or 2D ferro-type.
The former is plausible, and the latter should be unlikely due to antiferromagnetic structure of Ba$_{1-x} $Cs$_x$Fe$_2$Se$_3$.
In stripe-II magnetism region, ordered magnetic moments are smaller than that of parent compounds and thus fluctuating moment may be significant.
This part may contribute to the larger $\gamma$ value in the stripe-II region.
However, the parent compounds, which have larger moment size than intermediate compounds, do not show finite $\gamma$ values.
Therefore it would be difficult to conclude that the contribution of magnon is dominant.

Finally, for the case of electronic specific heat, electrons may have density of states at Fermi level, but random potential prevents their itinerancy.
In this possibility, the resistivity would follow the VRH model in low temperatures, and the magnetic moment would reduce because of its itinerancy.
This is consistent with the fact that similar fashion is seen in the $\gamma$ values and ($k_{\rm B}T_{\rm 0}$)$^{-1}$
The small moments of stripe-II magnetism is also consistent with this scenario.
These behaviors can be seen in insulating amorphous alloys, however theoretical model to account for $\gamma$ in insulating system is still controversial \cite{PhysRevB.61.15550}.
We suggest that this is the most plausible scenario among these possibilities.

The obtained results indicate the close relation between the crystal and magnetic structures.
In the $Pnma$ structure, the intra-ladder structure is distorted and rungs are slightly tilted from the parallel alignment along the rung direction [Fig. \ref{atom_lattice_a-d}(d)].
It is reported that BaFe$_2$Se$_3$ shows enhancements of distortion in intra-ladder structure below $T_{\rm N}$ $\sim 255$ K, and this indicates magnetoelastic coupling \cite{PhysRevB.85.064413, PhysRevB.84.180409}.
On the other hand, at $x = 0.05$, where only magnetic diffuse scattering was observed, there is no drastic intra-ladder distortion in measured temperature range [Fig. \ref{local_structure}(c), \ref{local_structure}(d)].
In addition, at $x = 0.15$ where the crystal structure has the $Cmcm$ space group, only weaker magnetic diffuse scattering than that of $x = 0.05$ was observed.
These facts suggest that no long range order of the block magnetism can exist in the $Cmcm$ structure, and that the block magnetism may be sensitive to the distortion of intra- and inter-ladder atom arrangement.

The stripe-I and II magnetism are more resilient than the block magnetism to the elemental substitution for $0.4 \leq x \leq 0.75$ region.
No structure transition was observed in $0.4 \leq x \leq 0.75$.
The stability of the  magnetic structure would be due to that of the crystal structure.

For $x = 0.75$, the local structure changes slightly at the transition temperatures [Fig. \ref{local_structure}(e), \ref{local_structure}(f)].
However the relation between the crystal and the magnetic structure is unclear at present.

Next we focus on the $x = 0.25$ sample, where no diffuse or long range magnetic scattering observed by powder neutron diffraction measurements down to 7 K.
In this case, moments would vanish or freeze without long range order nor short range correlation.
From susceptibility measurements, the moments seem to remain finite [Fig. \ref{chi_M_sum}(b)].
Hence, there must be fluctuating moment (or glassy component) even at the lowest temperature.
This compound has the competitions of block and stripe-II magnetism, as well as the $Pnma$ and $Cmcm$ structure.
The suppression of magnetism for $x = 0.25$ would be due to these instabilities.

We now compare obtained magnetic phase diagram (Fig. \ref{phase_diagram_latest_2}) with theoretical calculations based on a five-orbital Hubbard model in a finite size Fe-Se two-leg-ladder \cite{PhysRevB.87.024404}.
The calculations correspond to formal valences of Fe$^{2+}$ (correspond to BaFe$_2$Se$_3$, or $x = 0$) and Fe$^{2.25+}$ (Ba$_{0.5}$Cs$_{0.5}$Fe$_2$Se$_3$, or $x = 0.5 $).
They took account of the lattice distortion seen in the $Pnma$ space group as anisotropic hopping parameters, and they calculated the $x = 0.5$ sample as the $Pnma$ (experimental results indicate $x = 0.5 $ is $ Cmcm$).
Block and CX structure [corresponds to stripe-I and II magnetism, see Fig. 1 in Ref. \cite{PhysRevB.87.024404}] are stable in a robust region of their phase diagrams for Fe$^{2+}$ and Fe$^{2.25+}$, respectively.
Except for the orientation of spins, their results are basically consistent with our findings.
Interestingly, the stripe magnetism can be stabilized in the $Pnma $ space group.
Hence the theory suggests that the stability of stripe-I and II magnetism are due to not only that of crystal structure but also that of electronic structure.
Note that magnetic structures obtained by our study are realized in the realistic ratio $J_{\rm H}/U = 0.25$ and $U/W \sim 0.5$ [see in Ref. \cite{PhysRevB.87.024404}].
This implies Ba$_{1-x} $Cs$_x$Fe$_2$Se$_3$ would be "intermediate" coupling compounds \cite{dai2012magnetism}.

Pressure-induced superconductivity was reported in cuprate ladder compounds Sr$_{14-x}$Ca$_x$Cu$_{24}$O$_{41}$ \cite{doi:10.1143/JPSJ.65.2764, PhysRevLett.81.1090, PhysRevB.57.613}.
This compound has Cu$_2$O$_3$ combined two-leg-ladder structure, and is insulator.
Their resistivity becomes lower with increasing pressure and show superconductivity at 3 GPa.
Moreover, the suppression of magnetism provides superconductivity in 2D iron-based SCs, and we discovered that the magnetism of Ba$_{0.75}$Cs$_{0.25}$Fe$_2$Se$_3$ ($x = 0.25$) is completely suppressed.
Therefore applying pressure for this composition may be an intriguing way to pursue superconductivity in future.
Pressure-effect study is now in progress.

\section{\label{conclusion}CONCLUSION}
We performed a comprehensive study on the interplay of crystal structure, bulk properties, and magnetism for Ba$_{1-x} $Cs$_x$Fe$_2$Se$_3$ by electrical resistivity, specific heat, magnetic susceptibility, X-ray diffraction and powder neutron diffraction measurements.
The space group changes from $Pnma$ to $Cmcm$ between $x = 0.05$ and 0.15, where block magnetism is drastically suppressed concomitantly. 
New type of magnetic structure, stripe-II, appeared where $T$-linear term of $C_p$ is finite.
For $x = 0.75$ we observed successive phase transitions from stripe-II to stripe-I with a slight structure distortion.
Typical magnetic moment sizes in Ba$_{1-x} $Cs$_x$Fe$_2$Se$_3$ was determined as: stripe-II magnetism $\sim 0.5 \mu_{{\rm B}}$; stripe-I $ \sim 2\mu_{{\rm B}}$; and block magnetism $ \lesssim 3\mu_{{\rm B}}$.
They cover the moment size range seen in the all iron-based SCs: single-stripe magnetism $< 1\mu_{{\rm B}}$; double-stripe $\sim $2$\mu_{{\rm B}}$; and block magnetism $\sim 3\mu_{{\rm B}}$.
Notably, there is no diffuse nor resolution-limited magnetic scattering in powder neutron diffraction pattern down to 7 K for $x = 0.25$, where competition of the block and the stripe-II magnetism and of the $Pnma$ and $Cmcm$ structure is expected.

\begin{acknowledgements}
The work at ISSP and IMRAM is supported by KAKENHI (26800175, 23340097, 23244068, and 24224009).  The work at IMRAM is also supported by Cooperative Research Program of "Network Joint Research Center for Materials and Devices", and "Nano-Macro Materials, Devices and System Research Alliance". The research at Oak Ridge National Laboratory’s High Flux Isotope Reactor was sponsored by the Scientific User Facilities, Office of Basic Energy Sciences, U.S. Department of Energy.
\end{acknowledgements}

\bibliography{ref.bib}

\begin{thebibliography}{42}
\bibitem [{mag()}]{magnetization_impurity}%
  \BibitemOpen
  \href@noop {} {}\bibinfo {note} {To estimate the effect of ferrimagnetism of
  Fe$_7$Se$_8$, we took hysteresis loops of the samples and pelleted
  Fe$_7$Se$_8$ between $H $ = -10000 and 10000 Oe at room temperature, and
  compared with that of Ba$_{1-x} $Cs$_x$Fe$_2$Se$_3$. We had annealed
  Fe$_7$Se$_8$ using the same sequences as for Ba$_{1-x} $Cs$_x$Fe$_2$Se$_3$
  before measurements and measure it, since Fe$_7$Se$_8$ has two
  superstructures 4$c$ and 3$c$ \cite{doi:10.1143/JPSJ.16.1162}.}\BibitemShut
  {Stop}%
\end{thebibliography}

\end{document}